\documentclass[aps,prl,twocolumn,groupedaddress]{revtex4-1}
\usepackage[english]{babel}
\usepackage[utf8]{inputenc}
\usepackage[dvips]{graphicx}
\usepackage[active]{srcltx}

\begin{document}

\title{Polytypic nanowhiskers: electronic properties in the vicinity of the band edges}
\author{Faria Junior, P. E. and Sipahi, G. M.}
\affiliation{Instituto de Física de São Carlos, USP, São Carlos, SP, Brazil}
\date{\today}


\begin{abstract}
The increasing interest of nanowhiskers for technological applications has led to the 
observation of the zinc-blend/wurtzite polytypism. Polytypic nanowhiskers could also play, 
by their characteristics, an important role on the design of optical and electronic devices. 
In this work we propose a theoretical model to calculate the electronic properties of 
polytypic zinc-blend/wurtzite structure in the vicinity of the band edges. Our formulation 
is based on the $\vec{k} \cdot \vec{p}$ method connecting the irreducible 
representations in the interface of the two different crystalline phases by group 
theory arguments. Analyzing the composition of the states in the $\Gamma$ 
point and the overlap integrals of the envelope functions we predict energy 
transitions that agree with experimental photoluminescence spectra.
\end{abstract}

\maketitle


\section{Introduction}

Recently an increasing number of papers have reported high quality and control growth of 
nanowires and nanowhiskers (NWs) of III-V compounds such as InP, InAs, GaAs and  
GaP \cite{am-21-3654,jcg-298-635,naturemat-5-574,naturenano-4-50,nl-7-1500,nl-10-1699,nanoIEEE-6-384}. 
Nanoelectronics, nanophotonics and solar cells are some of the technological applications for these 
nanostructures like described in the references above.

Vapor-liquid-solid (VLS) method \cite{apl-4-89} was used, in 1964, by Wagner and Ellis to grew 
a silicon whisker. Since then, other methods like molecular beam epitaxy (MBE) 
and metal organic vapor phase epitaxy (MOVPE) have also been applied to the growth of NWs. 
These techniques allow the fabrication of homo and heterostructures in both axial and 
radial directions opening new opportunities to engineering novel device applications.

A heterostructure is composed by assembling different materials side-by-side along a 
growth direction. The homostructure is a similar concept, however, it is not the 
kind of material that is different but its crystalline structure. This effect is 
also known as polytypism and have been observed in NWs structures of several III-V compounds.

Under standard temperature and pressure conditions, the III-V compounds have as the bulk stable phase the 
zinc-blend (ZB) structure, exception made for the nitrides that have the wurtzite (WZ) structure as the 
most stable phase. However, reducing the dimensions to the nanoscale level such as in these NWs, the WZ 
phase has been observed for non-nitride III-V compounds. The natural explanation to this behavior is the 
small formation energy difference between the two crystal structures.


The phase alternation creates ZB and WZ regions along the growth direction affecting the 
NWs optical and electronic properties, generating new forms of band offsets and electronic minibands 
that can be explored to produce new kinds of applications.

The WZ/ZB polytypism in III-V compounds can make a type II homostructure characterized 
by a staggered band-edge lineup, i. e., the charge carriers are spatially 
separated: the electrons will be concentrated in the ZB region while the holes will 
be in the WZ region. A summary of NWs growth, properties and applications was made by 
Dubrovskii {\it et al.}~\cite{semiconductors-43-1539}.

From the theoretical point of view, the literature about these III-V NW structures, 
is focused in growth stability, surface energies and formation of polytypism 
\cite{jap-104-093520,jcg-301-302-862,jjap-45-L275, nano-21-285204,pss-52-1531, 
semiconductors-44-127,tpl-35-99}. Only recently theoretical calculations 
of the electronic properties in the band edges have been done for such structures 
~\cite{prb-82-125327}. None has been done in the vicinity of the band edges.

The polytypism was investigated in a paper of Murayama and Nakayama \cite{prb-49-4710} 
by \textit{ab initio} calculations of the band offsets in WZ/ZB interfaces. A useful 
schematics of the correspondence between the symmetry of the states of the WZ and ZB 
structures has been presented in the same article.

The aim of this study is to propose a theoretical model based on group theory arguments 
and the $\vec{k} \cdot \vec{p}$ method to calculate the electronic band structure in the vicinity
of the band edges and envelope functions of a WZ/ZB/WZ homostructure for indium phosphide (InP). 
The effective mass parameters for the WZ bulk phase of InP were extracted from a paper by Pryor 
and De \cite{prb-81-155210} that predicted the band structures of some III-V bulk compounds in 
the WZ phase (phosphides, arsenides and antimonides). The ZB parameters are extracted from
Vurgaftman {\it et al.}~\cite{jap-89-5815}.

The formulation of a theoretical model for this new kind of nanostructures can 
lead to band gap engineering and novel electronic a and also work as a guide 
for the crystal growth community.


\section{Theoretical model}

The existence two different crystal phases in a single whisker, can be explained by a 
simple joint analysis of the ZB and WZ crystal structures. Looking into ZB
along the [111] direction and to WZ along the [0001] direction,
one can found a noticeable similarity. Both can be described as stacked
hexagonal layers. In fact, the ZB structure has three layers in the
stacking sequence (ABCABC) while the WZ structure has only two (ABAB).
So, a stacking fault in WZ becomes a single segment of ZB and two
sequential twin planes in ZB form a single segment of WZ \cite{naturenano-4-50}.

\begin{figure}[t]
\includegraphics{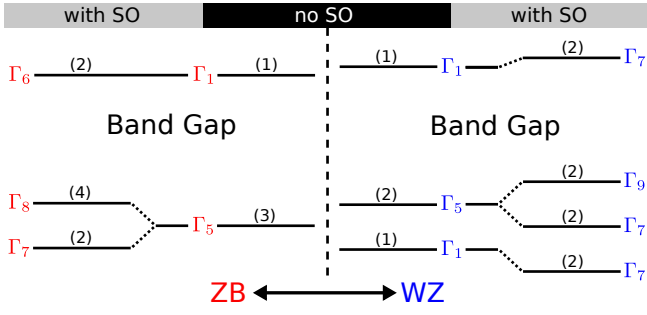}
\caption{$\Gamma$ point symmetry for the two crystal structures. The numbers in parentheses 
are the degeneracy of the irreducible representation. This figure was adapted form \cite{prb-49-4710}.}
\label{fig:gamma_symmetry}
\end{figure}

An important difference between the two structures is the crystal
field energy present in the WZ structure that defines the top of the
valence band, neglecting spin, as a composition of two different bands, 
a non-degenerate and a bidegenerate bands. In ZB, the top of valence band 
is composed by a three-degenerate band. This difference in the energy bands 
between the two structures can be seen schematically for the $\Gamma$ point
in figure \ref{fig:gamma_symmetry}, with and without the spin-orbit interaction.

To calculate the band structure of a polytypic NW based in the $\vec{k}\cdot\vec{p}$
matrix formalism, it is necessary to understand the differences between
the two crystal structures in the reciprocal lattice or, more precisely,
what happens with the high-symmetry points in the ZB first Brillouin
zone when we are dealing with the [111] growth direction and how
they can be related to the WZ ones.

\begin{figure}[t]
\includegraphics{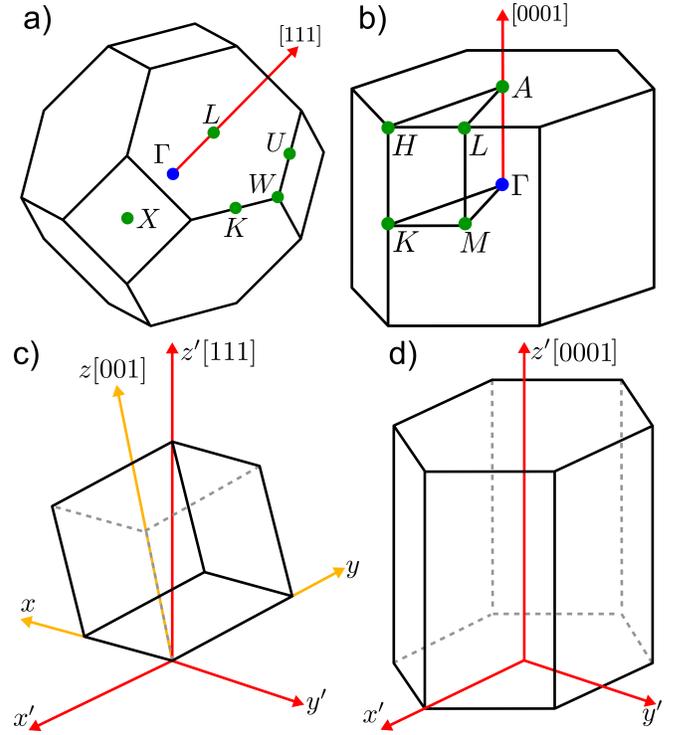}
\caption{(a) ZB and (b) WZ first Brillouin zones. The arrows (red online) represent 
the growth directions. (c) ZB conventional unit cell with two different coordinate systems. 
(d) WZ conventional unit cell with its common coordinate system. The [111] 
growth direction for ZB structure passes along the main diagonal 
of the cube and is represented in the primed coordinate system.}
\label{fig:BZ_coordinates}
\end{figure}

Looking through the [111] axis in figure \ref{fig:BZ_coordinates}a one can
see that the $L$ point of ZB structure is projected in the $\Gamma$ point. Since the number 
of atoms in the primitive unit cell for WZ is twice the number in ZB, there 
are more states in the $\Gamma$ point of WZ. This correspondence is done by 
assigning $\Gamma$ and $L$ points of the ZB to the $\Gamma$ point of WZ. This 
compatibility can be seen in the energy diagram presented by Murayama and 
Nakayama \cite{prb-49-4710}, that relates the irreducible representations of the 
ZB $\Gamma$ and $L$ points to the WZ $\Gamma$ point. This energy diagram provides 
the essential insight for our method since we are going to relate $\vec{k}\cdot\vec{p}$
matrices developed with irreducible representations belonging to different
crystal structures.

The usual $\vec{k}\cdot\vec{p}$ matrix formulation is constructed
in the vicinity of the $\Gamma$ point for ZB and WZ structures \cite{kane-book,prb-54-2491} 
considering the first three valence bands and the first
conduction band, represented by a $4\times4$ matrix. When the spin-orbit
interaction is included this number is multiplied by 2, resulting
in a $8\times8$ matrix. These bands are exactly the ones presented
in figure \ref{fig:gamma_symmetry}, and represent a smaller set of the states first presented
by Murayama and Nakamura \cite{prb-49-4710}. Analyzing this, one may notice
that, in the present method the irreducible representations of ZB [111]
belong only to the $\Gamma$ point and no irreducible representation of the
$L$ point will appear in the formulation.

Another subject, asides the symmetry correspondence, must be pointed
out: the $\vec{k}\cdot\vec{p}$ matrix for the ZB structure is developed
for the cubic main axes, which are the [100], [010] or [001]
growth direction. This implies that the definition of the $x,y,z$ coordinate
system defines the matrix representations of the symmetry operations
used to calculate the matrix elements in the $\vec{k}\cdot\vec{p}$ formulation.
Changing the coordinate system to $x^{\prime},y^{\prime},z^{\prime}$,
which is the coordinate system of the WZ Hamiltonian, will enable
us to relate the ZB and WZ matrices.

We can either recalculate the matrix representations of the symmetry
operations and develop the ZB $\vec{k}\cdot\vec{p}$ matrix or use
the rotation matrices presented in the paper of Park and Chuang \cite{jap-87-353}
to transform the ZB $\vec{k}\cdot\vec{p}$ Hamiltonian in the $x,y,z$
to the $x^{\prime},y^{\prime},z^{\prime}$ coordinate system.

The basis, $\left\{ \left|c_{l}\right\rangle \right\}$, in the prime coordinate system 
is defined as (from this point on, the prime will be dropped out of the notation):

\begin{eqnarray}
\left|c_{1}\right\rangle & = & -\frac{1}{\sqrt{2}}\left|(X+iY)\uparrow\right\rangle \nonumber \\
\left|c_{2}\right\rangle & = & \frac{1}{\sqrt{2}}\left|(X-iY)\uparrow\right\rangle \nonumber \\
\left|c_{3}\right\rangle & = & \left|Z\uparrow\right\rangle \nonumber \\
\left|c_{4}\right\rangle & = & \frac{1}{\sqrt{2}}\left|(X-iY)\downarrow\right\rangle \nonumber \\
\left|c_{5}\right\rangle & = & -\frac{1}{\sqrt{2}}\left|(X+iY)\downarrow\right\rangle \nonumber \\
\left|c_{6}\right\rangle & = & \left|Z\downarrow\right\rangle \nonumber \\
\left|c_{7}\right\rangle & = & i\left|S\uparrow\right\rangle \nonumber \\
\left|c_{8}\right\rangle & = & i\left|S\downarrow\right\rangle
\end{eqnarray}

which is convenient to describe the WZ Hamiltonian.

Since we are not treating the interband interaction explicit in the
matrix, the conduction band have a parabolic form for both crystals
and is not affected by the rotation, reading as

\begin{equation}
E_{c}(\vec{k})=E_{g}+E_{VBM}+\frac{\hbar^{2}}{2m_{e}^{\parallel}}k_{z}^{2}+\frac{\hbar^{2}}{2m_{e}^{\perp}}\left(k_{x}^{2}+k_{y}^{2}\right)
\end{equation}

where $E_{g}$ is the gap energy, $E_{VBM}$ is the energy of the
top valence band and for the ZB structure the electron effective masses
are the same $m_{e}^{\parallel}=m_{e}^{\perp}$.

Applying the rotation matrices to the valence band ZB Hamiltonian
we obtain a common matrix for the two structures:

\begin{equation}
H_{V}=\left[\begin{array}{cccccc}
F & -K^{*} & -H^{*} & 0 & 0 & 0\\
-K & G & H & 0 & 0 & \Delta\\
-H & H^{*} & \lambda & 0 & \Delta & 0\\
0 & 0 & 0 & F & -K & H\\
0 & 0 & \Delta & -K^{*} & G & -H^{*}\\
0 & \Delta & 0 & H^{*} & -H & \lambda\end{array}\right]
\end{equation}

The matrix terms are defined as:

\begin{eqnarray}
F & = & \Delta_{1}+\Delta_{2}+\lambda+\theta \nonumber \\
G & = & \Delta_{1}-\Delta_{2}+\lambda+\theta \nonumber \\
\lambda & = & A_{1}k_{z}^{2}+A_{2}\left(k_{x}^{2}+k_{y}^{2}\right) \nonumber \\
\theta & = &A_{3}k_{z}^{2}+A_{4}\left(k_{x}^{2}+k_{y}^{2}\right) \nonumber \\
K & = & A_{5}k_{+}^{2}+2\sqrt{2}A_{z}k_{-}k_{z} \nonumber \\
H & = & A_{6}k_{+}k_{z}+A_{z}k_{-}^{2} \nonumber \\
\Delta & = & \sqrt{2}\Delta_{3} 
\end{eqnarray}

where $k_{\pm} = k_x \pm i k_y$.

The only difference from the usual WZ formulation is the inclusion of a new parameter,
$A_{z}$, that takes into account the reorientation of the ZB growth axis.

The $A_{i}(i=1,...,6,z),$ and $\Delta_{i}(i=1,2,3)$ WZ parameters
can be related to the $\gamma_{i}(i=1,2,3)$ and $\Delta_{SO}$ ZB
parameters comparing the ZB rotated matrix to the WZ matrix.

In this work we will assume the WZ as ideal and therefore the lattice parameters
will considered the same as ZB. As a consequence, strain effects will not be
taken into account.


\section{Results and discussion}

We first apply our model for bulk ZB and WZ InP to check the reliability
of the matrix for both crystal structures. This initial investigation is
useful to fit the effective mass parameters and to understand the
valence band states in the band edge since now we are dealing with
different basis states for the ZB matrix.

\begin{figure}[t]
\includegraphics{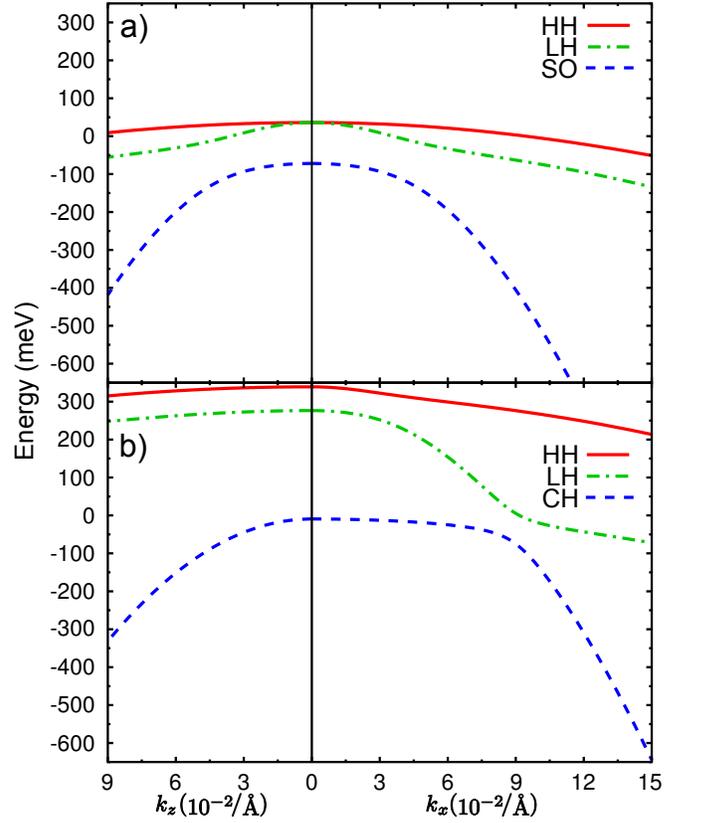}
\caption{(a) Bulk ZB valence band structure along the [111]
direction. (b) Bulk WZ valence band structure reproducing results
in ~\cite{prb-81-155210} along the [0001] growth
direction. Both valence band structures were plotted for 20\% of the
first Brillouin Zone in $k_{x}$ and $k_{z}$ directions. Without
spin-orbit the $(x,y,z)$ coordinates belong to the same irreducible
representation for ZB structure and then we can see that the band
structure is isotropic in $k_{x}$ and $k_{z}$ directions. In WZ
structure $(x,y)$ and $(z)$ coordinates belong to different irreducible
representations making the band structure anisotropic in $k_{x}$
and $k_{z}$ directions.}
\label{fig:bulks_vb}
\end{figure}

The ZB parameters were obtained from Ref.~\cite{jap-89-5815}
and the WZ parameters were derived using the effective masses
presented in Ref.~\cite{prb-81-155210}. Since in our model 
there is no explicit interband interaction only the valence band is 
presented in figure \ref{fig:bulks_vb} for the bulk form of both crystal structures.

The usual identification of the bands were used for ZB while in WZ
we had to analyze the composition of the states in the $\Gamma$ point.
Table \ref{tab:gamma_point_energies} summarizes the energies and states compositions in the $\Gamma$
point that helps us to identify the WZ states.

\begin{center}
\begin{table}
\begin{tabular}{ccc}
\hline 
Energy & Set of states 1 & Set of states 2\tabularnewline
\hline 
\hline 
$E_{1}$ & $\left|c_{1}\right\rangle $ & $\left|c_{4}\right\rangle $\tabularnewline
$E_{2}$ & $\alpha\left|c_{2}\right\rangle +\beta\left|c_{6}\right\rangle $ & $\beta\left|c_{3}\right\rangle +\alpha\left|c_{5}\right\rangle $\tabularnewline
$E_{3}$ & $\beta\left|c_{2}\right\rangle -\alpha\left|c_{6}\right\rangle $ & $-\alpha\left|c_{3}\right\rangle +\beta\left|c_{5}\right\rangle $\tabularnewline
\hline
\end{tabular}
\caption{The valence band energies and the respective eigenstates at the $\Gamma$ point.}
\label{tab:gamma_point_energies}
\end{table}
\end{center}

The energies and the $\alpha$ and $\beta$ coefficients are given by \cite{prb-54-2491}:

\begin{equation}
E_{1} = \Delta_{1}+\Delta_{2}
\end{equation}

\begin{equation}
E_{2} = \frac{\Delta_{1}-\Delta_{2}}{2}+\sqrt{\left(\frac{\Delta_{1}-\Delta_{2}}{2}\right)^{2}+2\Delta_{3}^{2}}
\end{equation}

\begin{equation}
E_{3} = \frac{\Delta_{1}-\Delta_{2}}{2}-\sqrt{\left(\frac{\Delta_{1}-\Delta_{2}}{2}\right)^{2}+2\Delta_{3}^{2}}
\end{equation}

\begin{equation}
\alpha = \frac{E_{2}}{\sqrt{E_{2}^{2}+2\Delta_{3}^{2}}}
\end{equation}

\begin{equation}
\beta = \frac{\sqrt{2}\Delta_{3}}{\sqrt{E_{2}^{2}+2\Delta_{3}^{2}}}
\end{equation}

The states identification were made as described in Ref.~\cite{apl-68-1657}:
HH (heavy hole) states are composed only by $\left|c_{1}\right\rangle $
or $\left|c_{4}\right\rangle $, LH (light hole) states are composed
mainly of $\left|c_{2}\right\rangle $ or $\left|c_{5}\right\rangle $
and CH (crystal-field split-off hole) are composed mainly of $\left|c_{3}\right\rangle $
or $\left|c_{6}\right\rangle $. 
According to the identification of the energies presented in table \ref{tab:gamma_point_energies},
$E_{1}$ is the HH state and, if $\alpha>\beta$, $E_{2}$ is the LH state and $E_{3}$ is the CH state or
$E_{2}$ is the CH state and $E_{3}$ is the LH state otherwise.
The valence band ordering in the $\Gamma$ point, which depends on the values of $\Delta_1$, 
$\Delta_2$ and $\Delta_3$, and the value of the coefficients
for both crystal structures can be found in table \ref{tab:alpha_beta_order}.

\begin{center}
\begin{table}
\begin{tabular}{cccc}
\hline 
 & Energies & $\alpha$ & $\beta$\tabularnewline
\hline
\hline  
ZB & $E_{1}=E_{2}>E_{3}$ & 0.57735 & 0.81650\tabularnewline
WZ & $E_{1}>E_{2}>E_{3}$ & 0.98345 & 0.18116\tabularnewline
\hline  
\end{tabular}
\caption{The energy ordering and the $\alpha$ and $\beta$ coefficients.
The knowledge of these coefficients enable the identification of the states.}
\label{tab:alpha_beta_order}
\end{table}
\end{center}

Since in our model we are using the WZ basis to describe the ZB structure
it is useful to think that ZB is a WZ structure with zero crystal-field
energy. Noting this feature we can identify the ZB valence band states
in the $\Gamma$ point using the WZ identification. This helps us
to understand the band edge in the WZ/ZB interface in a quantum well
structure. Table \ref{tab:WZ_ZB_identification} relates the two different identifications.

\begin{center}
\begin{table}
\begin{tabular}{cc}
\hline 
ZB identification & WZ identification\tabularnewline
\hline 
\hline 
HH & HH\tabularnewline
LH & CH\tabularnewline
SO & LH\tabularnewline
\hline
\end{tabular}
\caption{The WZ identification for the ZB states.}
\label{tab:WZ_ZB_identification}
\end{table}
\end{center}

As the dimensions of NWs are big enough to avoid lateral confinement, 
this structure can be considered as a 2-dimensional system, like a 
quantum well or a superlattice. In this context, we simulated a single 
WZ/ZB/WZ polytypic quantum well with $160\textrm{\AA}$ well width and 
$320\textrm{\AA}$ barrier width so that the whole system has $480\textrm{\AA}$.

\begin{figure}[t]
\includegraphics{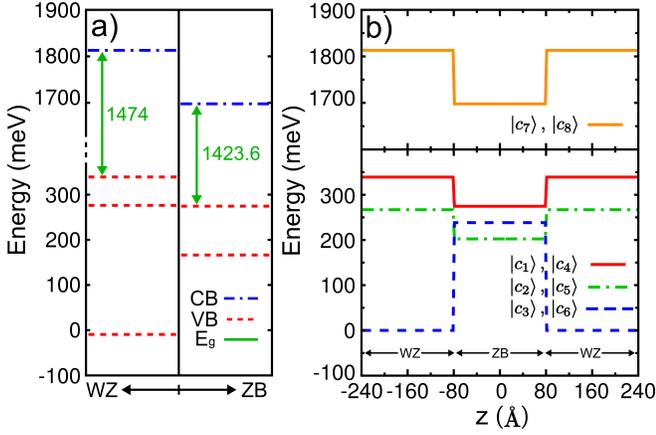}
\caption{(a) The band edge energy in the WZ/ZB interface relating the WZ and ZB states. 
(b) The diagonal potential profile of the total Hamiltonian along the growth direction.}
\label{fig:interface_diag_profile}
\end{figure}

Using the band mismatch value given in Ref.~\cite{prb-81-155210} it is possible
to construct the quantum well profile. To understand the confinement
profile we plotted, in figure \ref{fig:interface_diag_profile}a, the band edge 
energy in the WZ/ZB interface. Since the composition of the states LH and CH is different
in both crystal structures we did not think it was convenient to draw
a line connecting the energies. We believe that there's a mix of these
values to form the effective potential. However, we can note that
the conduction band and the first valence band form a type-II homostructure.

It is also convenient to plot the diagonal terms of the total Hamiltonian
to check the variation of the parameters along the growth direction,
which can be found in figure \ref{fig:interface_diag_profile}b. Although the nondiagonal 
terms of the matrix will mix the three profiles we can expect that the confinement 
of the electrons will be in the ZB region while the holes will be in the WZ region 
belonging in their most to the HH state.

\begin{figure}[t]
\includegraphics{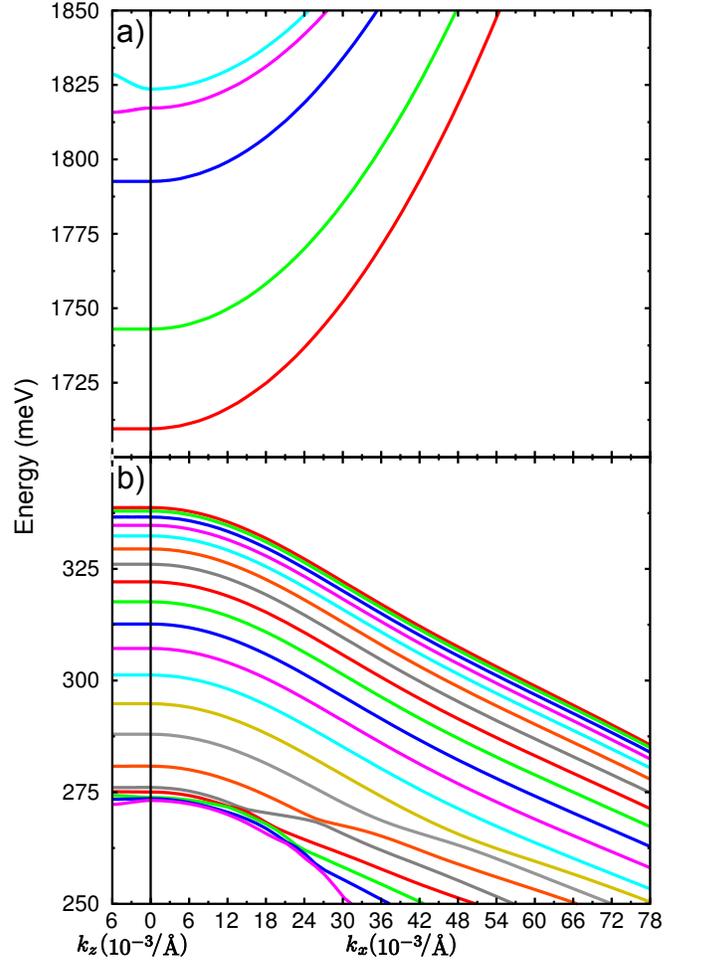}
\caption{Conduction (a) and valence (b) band structure of the single
WZ/ZB/WZ polytypic quantum well. From bottom to top the conduction
band states are: EL1 - EL5. From top to bottom the valence band states
are: HH1 - HH15, LH1, LH2, HH16, LH3, HH17.}
\label{fig:qw_bs}
\end{figure}

The band structure of the quantum well plotted up to 10\% of the $k_{x}$
direction and 100\% of the $k_{z}$ direction is shown in figure \ref{fig:qw_bs}.
We can observe that the bound states in the valence band are mainly
HH states, which means they are composed entirely by the $\left|c_{1}\right\rangle $
and $\left|c_{4}\right\rangle $ states of the basis that have only
$\left|X\right\rangle $ and $\left|Y\right\rangle $ components.
This feature directly reflects in the luminescence properties of the
system: the light will be almost fully polarized in the $xy$ plane, i. e., transversal
to the growth axis of the polytypic quantum well. The LH states have less than 5\% of 
$\left|Z\right\rangle$ component.

\begin{figure}[t]
\includegraphics{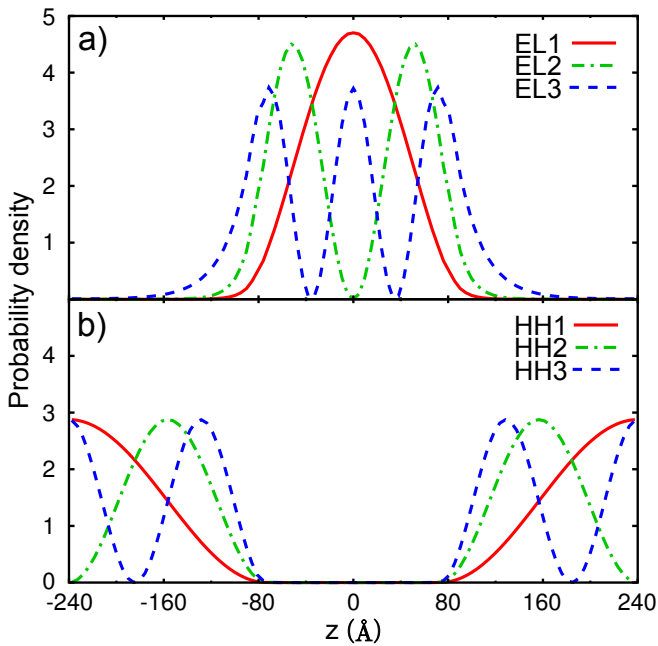}
\caption{The probability densities of the electrons (a) and holes
(b). We can note the spatial separation of the carriers.}
\label{fig:prob_dens}
\end{figure}

The probability densities calculated from the envelope functions for 
the $\Gamma$ point of the EL1, EL2, EL3, HH1, HH2 and HH3 bands are 
shown in figure \ref{fig:prob_dens}. These results agrees with the 
experimental data suggesting a spatial separation of the carriers, 
due to the fact that is more likely to find holes in the WZ region 
and electrons in the ZB region.

In a quantum well the selection rules for dipole optical transition
are given by the matrix element $M$ \cite{markfox}. 

\begin{equation}
M = \left\langle E\left|\vec{r}\cdot\hat{\varepsilon}\right|H\right\rangle = M_{c}M_{g}
\end{equation}

This matrix element can be divided in two parts:
$M_{c}$ and $M_{g}$. $M_{c}$ is the matrix element between the
electron and hole states in the basis $\left\{ \left|c_{l}\right\rangle \right\}$
that depends on the polarization of the light and $M_{g}$ is the
overlap integral between the electron and hole envelope functions.

\begin{equation}
M_{c} = \left\langle c_{e}\left|\vec{r}\cdot\hat{\varepsilon}\right|c_{h}\right\rangle
\end{equation}

\begin{equation}
M_{g} = \int_{-\infty}^{\infty}g_{e}^{*}(z)\, g_{h}(z)\, dz
\end{equation}

The $M_{c}$ matrix elements are known from symmetry properties of
the $\Gamma$ point and the non-vanishing elements are
$\left\langle S\left|x\right|X\right\rangle$, $\left\langle S\left|y\right|Y\right\rangle$
and $\left\langle S\left|z\right|Z\right\rangle$. We just have
to calculate the overlap integral $M_{g}$. Since the potential profile is
even, the envelope functions have well defined parities, they are either
even or odd and we can expect to have null overlap integrals between
conduction and valence states. The non-zero overlap integrals and the difference energy in 
the $\Gamma$ point are presented in table \ref{tab:ovlp_integrals}.

\begin{center}
\begin{table}
\begin{tabular}{ccc}
\hline 
 & $M_{g}$ (\%) & Energy difference (meV)\tabularnewline
\hline 
\hline 
EL1-HH1 & $\sim$2.6 & 1370.76\tabularnewline
EL1-HH3 & $\sim$6.4 & 1372.89\tabularnewline
EL2-HH2 & $\sim$12 & 1405.06\tabularnewline
EL3-HH1 & $\sim$23 & 1453.86\tabularnewline
EL3-HH3 & $\sim$33.5 & 1455.99\tabularnewline
\hline
\end{tabular}
\caption{Finite overlap integrals ($M_{g}$) and energy difference between
some conduction and valence band states at the $\Gamma$ point. We
can note that overlap is bigger for excited states.}
\label{tab:ovlp_integrals}
\end{table}
\end{center}

Although we have achieved reliable results with the model considered here 
for InP, we believe that improvements should be made to treat other III-V 
compounds. In systems where the lower conduction band has symmetry $\Gamma_7$ 
in the WZ phase and the spin-orbit splitting is bigger or equivalent to the 
gap energy (InAs and InSb \cite{prb-81-155210}) the full $8\times8$ $\vec{k}\cdot\vec{p}$ 
method should be considered with the inclusion of the Kane interband momentum matrix 
elements ($P$) and also Kane quadratic terms ($B$). In systems with $\Gamma_8$ symmetry 
in the lower conduction band for WZ phase, like AlAs or GaAs \cite{prb-81-155210}, 
the full symmetry analysis should be redone.


\section{Conclusions}

In summary, we have presented a theoretical method based in group theory 
arguments and the $\vec{k}\cdot\vec{p}$ method to calculate the electronic 
properties of InP polytypic structures in the vicinity of the band edges.

Understanding the behavior of the band structure of the calculated system 
we could predict the trends in the luminescence energy transitions.
The preliminary results are in good agreement with the experimental data 
available so far such as carriers' spatial separation and light polarization 
in luminescence spectra.

With the proposed model, it will be possible to predict the optical 
properties of selected structures and to design systems that best fit new 
device applications.


\section{Acknowledgements}

We like to thank the Brazilian funding agencies CAPES and CNPq for the 
financial support.


\bibliography{artigos}

\end{document}